\documentclass{osa-article}

\journal{oe}

\usepackage{hyperref}
\usepackage{cite}
\usepackage{upgreek}
\newcommand{\st}[1]{} \renewcommand{\emph}[1]{#1}

\begin{document}

\title{Optimal binary gratings for multi-wavelength magneto-optical traps}

\author{Oliver S. Burrow,\authormark{1,*} 
Robert J.\ Fasano,\authormark{2,3} 
Wesley Brand,\authormark{2,3} 
Michael W.\ Wright,\authormark{1}   
Wenbo Li,\authormark{1} Andrew D.\ Ludlow,\authormark{2,3} Erling~Riis,\authormark{1} Paul F.\ Griffin,\authormark{1} and Aidan S. Arnold\authormark{1,$\dagger$}}
\address{\authormark{1}Department of Physics, SUPA, University of Strathclyde, Glasgow, G4 0NG, United Kingdom}
\address{\authormark{2}National Institute of Standards and Technology, 325 Broadway, Boulder, Colorado 80305, USA}
\address{\authormark{3}University of Colorado, Department of Physics, Boulder, Colorado 80309, USA}

\email{\authormark{*}oliver.burrow@strath.ac.uk}
\email{\authormark{$\dagger$}aidan.arnold@strath.ac.uk}

\homepage{https://eqop.phys.strath.ac.uk/atom-optics/grating-mots}

\begin{abstract}
Grating magneto-optical traps  are an enabling quantum technology for portable metrological devices with ultracold atoms. \st{Gratings are chromatic, with} \emph{However}, beam diffraction efficiency and angle \emph{are} affected by wavelength, creating a \emph{single-optic design} challenge for laser cooling in two stages at two distinct wavelengths -- as commonly used for loading e.g.\ Sr or Yb  atoms into optical lattice or tweezer clocks. Here, we optically \emph{characterize} a wide variety of binary gratings at different wavelengths to find a simple empirical fit to experimental grating diffraction efficiency data in terms of dimensionless etch depth and period for various duty cycles. The model avoids complex 3D  light-grating surface \emph{calculations}, yet still yields results accurate to a few percent across a broad range of parameters. Gratings \emph{optimized} for two (or more) wavelengths can now be designed in an informed manner suitable for a wide class of atomic species enabling advanced quantum technologies.
\end{abstract}

\section{Introduction}

Laser cooled atoms are required for a wide range of quantum technologies, and there is growing demand to create portable, compact and robust devices capable of leaving \emph{controlled laboratory} environments \cite{bidel2018absolute,liu_-orbit_2018,becker_space-borne_2018,Grotti2018,aveline_observation_2020,Takamoto2020,Little2021}.
The grating magneto-optical trap (GMOT) \cite{Vangeleyn2010, Nshii2013, Lee2013,  McGilligan2015,mcgilligan2017grating,Imhof2017,Bregazzi2021} simplifies laser cooling of thermal atomic \emph{vapors} or beams when compared to traditional methods, reducing the optical system requirements to a single input beam and a planar optic.  
This reduction in size and  complexity enables compact cold-atom sources \cite{Scherschligt2018,McGilligan2020APL,Burrow2021}. There are a number of advantages to tetrahedral pyramid\cite{Vangeleyn2009} and grating-based designs over standard single-input beam pyramidal geometries \cite{Lee1996,Pollock2009,Bodart2010,Wu2017}, \emph{namely: the mitigation of absorption-induced beam shadows particularly for larger MOTs; high optical access; and the optic can be both mass-produced and used ex-vacuo}. \emph{This has led to increasing interest in the GMOT technique for a variety of sensing and physics applications \cite{Elvin2019,Franssen2019,vanNinhuijs2019,Barker2019,Weiner2020,Gehl2021,Lee2021,McGehee2021,Sun2021,Seo2021,Chen2022,Barker2022,Duan2022,Lee2022}}. 

The laser wavelength is critical for laser cooling, and must correspond to a suitable transition in the atomic (or molecular \cite{Barry2014,Truppe2017,Wu2021}) species being cooled.  Gratings are chromatic by nature, with laser wavelength particularly affecting both diffraction angle and efficiency -- thereby affecting the function of a GMOT. 
GMOTs have been \emph{realized} for the atomic species Rb \cite{Nshii2013}, Li\cite{Barker2019} and Sr \cite{Sitaram2020,bondza22,elgee22} with grating optics designed specifically for these elements. Furthermore, Rb \emph{optimized} binary gratings \cite{Cotter2016,McGilligan2016} have been demonstrated to have suitable radiation pressure balance over a broad wavelength range covering all alkali metals (Fig.~6 in \cite{McGilligan2016}), and are particularly appropriate for e.g.\ dual-species potassium and rubidium experiments \cite{Abend2023}.

The broadband performance of the gratings can be exploited to design GMOT optics suitable for laser cooling at multiple wavelengths. 
For example, in Yb and Sr optical lattice clocks, large atom numbers are typically collected from a thermal source using the primary cooling transition, at a `blue' wavelength ($\lambda_\textrm{b}$) with a broad natural linewidth that connects singlet states $|g\rangle\leftrightarrow|e_\textrm{b}\rangle$.  Subsequently, in order to reach ultracold temperatures and higher densities for loading into an optical dipole trap or lattice, this is usually followed by cooling on a secondary transition, at a more `red' wavelength ($\lambda_\textrm{r}$) with a narrow linewidth that connects a singlet to a triplet state  $|g\rangle\leftrightarrow|e_\textrm{r}\rangle$.
Developing GMOT optics suitable for these species will \emph{aid further development of} compact, portable optical lattice clocks \emph{\cite{Brand2023,Grotti2018,Takamoto2020}}.

\begin{figure}[!b]
	\centering	\includegraphics[width=\columnwidth]{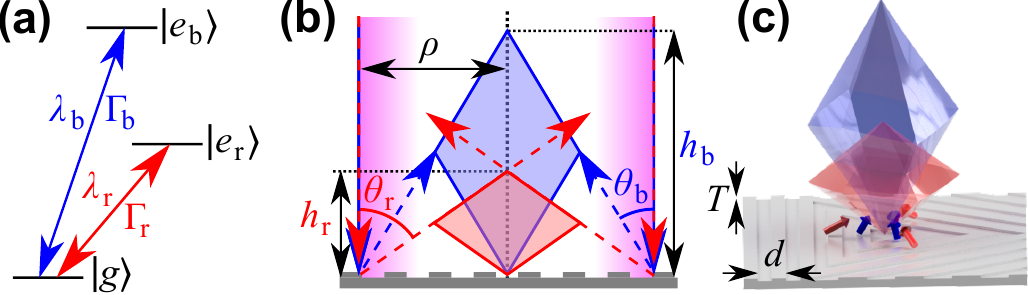}
	\caption{\label{Fig1} Simplified three-state atomic energy level diagram (a), with \emph{species-specific} `red' and `blue' transition\st{s with}  wavelengths and natural linewidths of $\lambda_\textrm{r,b}$ and $\Gamma_\textrm{r,b},$ respectively (Table~\ref{Table1}). The \emph{approximate 2D} schematic (b) illustrates how overlaid `red' and `blue' beams of radius $\rho$ propagate downwards onto a grating (gray), diffract at different first-order Bragg angles $\theta_\textrm{r,b}=\arcsin(\lambda_\textrm{r,b}/d)$, and form beam overlap volumes (shaded diamonds) with corresponding heights $h_\textrm{r,b}=\rho \cot \theta_\textrm{r,b}$, respectively. \emph{A more accurate 3D rendering of the overlap volumes is shown in (c)}. \st{MOTs can in principle form and be moved anywhere in the overlap zone corresponding to the wavelength used, via the central zero of the magnetic quadrupole field supplied by appropriate coils (not shown). Any `handover' between the MOTs (unaided by a transport trap) must occur in the intersection of the two overlap volumes.} The grating period is $d$ and etch depth $T$ which is the same nomenclature as  in \cite{McGilligan2016,Cotter2016}, except here the \emph{etched fraction of the period} is $p_\textrm{e}$ instead of $1-r$ to avoid confusion with the `red' r labels.}
\end{figure}

The level scheme and concept of the bi-chromatic grating MOT are shown in Fig.~\ref{Fig1} (a)-(c).  
A summary of some common laser-cooled elements with their `red' and `blue'  transitions can be seen in Table \ref{Table1}. 
The most noticeable chromatic grating effect is the different \emph{red (r) and blue (b)}  $m^\textrm{th}$-order diffraction angles $\theta_\textrm{r,b}$ as governed by Bragg's law $ \theta_\textrm{r,b}=\arcsin(m\lambda_\textrm{r,b}/d)$ for wavelengths $\lambda_\textrm{r,b}$ normally incident on a grating of period $d$.  Consequentially the capture volume height  $h_\textrm{b,r}=\rho \cot\theta_\textrm{b,r}$ is smaller  for the `red' MOT for the same  input beam radius $\rho$, as illustrated in Fig.~\ref{Fig1} (b,c). Here we constrain grating periods $\lambda_\textrm{r}<d<2\lambda_\textrm{b},$ to simplify the `blue' cooling with only first-order diffraction ($\theta_\textrm{b}>30^{\circ}$) and have `red' cooling at a real angle ($\theta_r<90^{\circ})$. 
Precluding second-order diffraction is not strictly necessary for MOT operation (see \cite{Sitaram2020}), but does mean that no other beam overlap regions and their radiation pressure balance need to be studied, or any transitions between these regions.

\begin{table}[!t]
\begin{tabular}{|l|l|l|l|l|l|l|l|l|l|}%l|
\hline
\multicolumn{1}{|l|}{}  
& \multicolumn{4}{c|}{Divalent atoms} & %\multicolumn{2}{c|}{Lanthanide}  &
\multicolumn{5}{c|}{Alkali metals} \\ \hline
Element & Ca & \textbf{Sr} & Yb & Tm & \textbf{Li} & Na & K & \textbf{Rb} & Cs \\ \hline
$|g\rangle$ & (4)$^1\textrm{S}_0$ & (5)$^1\textrm{S}_0$  & $^1\textrm{S}_0$ & $^2\textrm{F}_{7/2}$ & 2S$_{1/2}$ & 3S$_{1/2}$ & 4S$_{1/2}$ & 5S$_{1/2}$  & 6S$_{1/2}$ \\ \hline
$|e_\textrm{r}\rangle$ & $^3\textrm{P}_1$ & $^3\textrm{P}_1$ & $^3\textrm{P}_1$ & -- & 2P$_{3/2}$ & 3P$_{3/2}$ & 4P$_{3/2}$  & 5P$_{3/2}$ & 6P$_{3/2}$ \\ \hline
$|e_\textrm{b}\rangle$ & $^1\textrm{P}_1$ & $^1\textrm{P}_1$ &  $^1\textrm{P}_1$ & --  & 3P$_{3/2}$ & 4P$_{3/2}$ & 5P$_{3/2}$   & 6P$_{3/2}$ & 7P$_{3/2}$ \\ \hline
$\Gamma_\textrm{r}\,$(MHz) & 0.0004 & 0.007 & 0.183 & 0.345  & 5.9 & 9.8 & 6.0 & 6.1 & 5.2 \\ \hline
$\lambda_\textrm{r}\,$(nm) & 657 & \textbf{689}  & 556 & 531 & \textbf{671} & 589 & 767 & \textbf{780} & 852 \\ \hline 
$\lambda_\textrm{b}\,$(nm) & 423 & \textbf{461} & 399 & 411  & 323 & 330 & 405 & 420 & 455 \\ \hline
$\Gamma_\textrm{b}\,$(MHz) & 35 & 32 & 31 & 10  & 0.16 & 0.44 & 0.18 & 0.28 & 0.29 \\ \hline
$\lambda_\textrm{b}/\lambda_\textrm{r}$ & 0.644 & 0.669  & 0.718 & 0.774 & 0.481 & 0.560 & 0.528 & 0.538 & 0.534 \\ \hline
\end{tabular}
\caption{A summary of the `red' and `blue' cooling transition parameters (subscripts r and b, respectively) for a selection  of laser-cooled elements. The laser wavelength ($\lambda$) and transition linewidth ($\Gamma$) values are from the \href{https://www.nist.gov/pml/atomic-spectra-database}{NIST database}. GMOTs have already been realized for the atomic species and wavelengths in bold  \cite{Barker2019,Nshii2013,Sitaram2020,bondza22,elgee22}.} 
\label{Table1}
\end{table}

Alkaline earth and lanthanide\cite{Frisch2012,Ludlow2015,Golovizin2021} elements have a broad primary `blue' $|g\rangle\leftrightarrow|e_\textrm{b}\rangle$ and much narrower `red' $|g\rangle\leftrightarrow|e_\textrm{r}\rangle$ wavelength  cooling transition (Table~\ref{Table1}). 
Owing to different atomic structure \cite{Hall1989} these roles are reversed for alkali metals, with loading on the broad `red' transition and an option for  `blue' cooling  \cite{Duarte2011,Sebastian2014,Chen2016,McKay2011,Unnikrishnan2019} on a 20-40 times narrow transition, as first suggested in Ref.~\cite{Hall1989}. Again, the  narrower linewidth of the secondary cooling yields a lower temperature, and moreover the lower cross-section for reabsorption ($\propto \lambda^2$) means higher densities can also be achieved. High phase-space density can then be \emph{optimized}, particularly for species such as Li\cite{Duarte2011,Sebastian2014,Chen2016} and K\cite{McKay2011,Unnikrishnan2019} with closely-spaced hyperfine levels on the `red' line complicating direct sub-Doppler cooling, but also for Rb \cite{Das2023,Claussen}. For $^6$Li a blue MOT leads to 7 times more quantum degenerate atoms than even gray molasses \cite{Satter2018}. Aside from complexity there may be advantages to two-color cooling for other atomic species too, for higher-yield quantum degenerate gas or higher-accuracy quantum metrology experiments. 

We here consider microfabricated binary grating chips comprising $N$ sectors of one-dimensional gratings occupying equal angular regions of $2\pi/N\,$rad on the microfabricated grating surface. The use of two-dimensional gratings such as checkerboards \cite{Nshii2013} significantly improves overlap volume size and thereby MOT atom number. However, to eliminate higher-order diffraction one requires the more restrictive condition $\lambda_\textrm{b}/\lambda_\textrm{r}\geq 1/\sqrt{2}$ for 2D gratings (cf.\  $\lambda_\textrm{b}/\lambda_\textrm{r}\geq 1/2$ for 1D gratings), which eliminates most elements in Table~\ref{Table1} (and if $\theta_\textrm{b}$ is \emph{minimized} to $45^\circ$, then $\theta_\textrm{r}$ would be $80^\circ$ for Yb and $66^\circ$ for Tm).

Using the central zero of the magnetic quadrupole field supplied by appropriate coils (not shown in Fig.~\ref{Fig1} (b),(c)) MOTs can form and be moved anywhere in the overlap volume of all laser beams at the wavelength used.
Overlap volume mismatch between the `blue' and `red' wavelengths is a minor issue, which can be solved by ramping the MOT's magnetic quadrupole field centre when transferring from primary to secondary cooling transitions \cite{bondza22}, as long as MOT handover \emph{(if unaided by a transport trap)} occurs in the overlap volume intersection.
For alkali metals, where the primary cooling is a `red' wavelength, this problem can be mitigated by creating a smaller beam for the shorter wavelength to overlap the trap centres. There is overlap of the volumes regardless of beam size and ratio $\lambda_\textrm{b}/\lambda_\textrm{r}$, and since atoms are pre-cooled to a dense ball on the broad-linewidth transition, a smaller overlap volume on the narrow  transition is unlikely to be a problem anyway. 

\emph{The standard, albeit complex, approach to determine the behaviour of any specific grating is to solve polarisation-dependent Maxwell equations in 3D for a given light-surface boundary interaction comprising sub-wavelength periodic features \cite{MaxwellCode}.} % 
However, in the remainder of the paper we show that a simple empirical dimensionless fit suffices to \emph{accurately} describe experimental grating characteristics over a wide range of physical parameters. 
This model can then be used, in conjunction with general GMOT characteristics and metal coating reflectivity, to elucidate the grating properties for any combination of wavelengths and thus atomic or molecular species. \emph{The} specific examples of Sr and Yb \emph{are used} to illustrate the protocols, as they are increasingly valuable elements for ultra-precise atomic metrology.  

\section{Dimensionless empirical fit to experimental grating characteristics}
\label{results1}

Several hundred $2\,\textrm{mm}\times 2\,\textrm{mm}$ binary 1D gratings were fabricated using electron-beam lithography\cite{Cotter2016} varying the three physical grating parameters -- etch depth $T$, period $d$ and etched duty cycle fraction $p_\textrm{e}$ -- covering the ranges $105\,\textrm{nm}-195\,\textrm{nm}$, $600\,\textrm{nm}-4000\,\textrm{nm}$ and $40\%-85\%$, respectively.  Most gratings were coated with $100\,$nm or $200\,$nm Al, which have similar reflectivities \cite{McGilligan2016} and a few gratings had the same thicknesses of Au. All gratings were fully \emph{characterized}  using the method in Ref.~\cite{McGilligan2016} with %light 
wavelengths mainly targeting specific atomic species: i.e.\ $399\,\textrm{nm}$, $420\,\textrm{nm}$, $461\,\textrm{nm}$, $532\,\textrm{nm}$, $556\,\textrm{nm}$, $689\,\textrm{nm}$ and $780\,\textrm{nm}$. The diffractive properties of over 700 different grating-laser combinations were tested, particularly the first ($\eta_1$) and zeroth ($\eta_0$) order diffraction efficiencies for incident circularly \emph{polarized} light  \cite{McGilligan2016}.

In order to find an empirical model to fit our optical grating  \emph{characterizations}, \emph{they were} compared as a function of their dimensionless parameters: etch depth $T/\lambda$ and first-order diffraction angle $\theta=\arcsin(\lambda/d).$ In our prior work \cite{McGilligan2016}, the model used assumed constant total power in the combined first and zeroth diffracted orders, i.e.~constant $\eta_\textrm{t}=2 \eta_1+\eta_0$. Here, to get a better fit to \emph{the data, as described below, an} angular fit factor of the form $(1 - C \, \theta^2)$ is used, with $\theta$ here in degrees \emph{and $C$ a fit constant}.  
\emph{All data was corrected for less-than-unit reflectivity, i.e.\ all diffraction efficiencies were divided} by the reflectivity of the grating's metal coating at the measurement wavelength  \cite{weber2002handbook}.

\begin{table}[!b]
\centering
\begin{tabular}{|l||l|l|l|l|l||l|l|l|l|l|}%l|
\hline
$p_\textrm{e}$ & $A_1$ & $B_1$ & $C_1$ & $\phi_1$ & $\Delta\eta_1$ & $A_\textrm{t}$ & $B_\textrm{t}$ & $C_\textrm{t}$ & $\phi_\textrm{t}$ & $\Delta\eta_\textrm{t}$\\ \hline 
(45,55] & 22.6 & -18.1 & 74 & 1.37 & 4.1 & 86.9 & 11.0 & 14 & 0.79 & 6.6\\ \hline
(55,65] & 23.2 & -18.7 & 58 & 1.41 & 3.2 & 89.2 & 8.2 & 27 & 0.91 & 5.6\\ \hline
(65,75] & 25.9 & -12.4 & 55 & 1.44 & 4.2 & 88.7 & 9.4 & 28 & 1.14 & 6.5\\ \hline
\end{tabular}
\caption{The fitted surfaces for $\eta_1$ (left, subscript 1) and $\eta_\textrm{t}$ (right, subscript t) via Eq.~\ref{fiteq} for three etch duty ranges $p_\textrm{e}$ (in interval notation), with corresponding RMS fit errors $\Delta\eta_1$, $\Delta\eta_\textrm{t}$. The units of all quantities in the table are: $\%$ (for $p_\textrm{e},$ $A_i$, $B_i$, $\Delta\eta_i$, i.e.\ units matching $\eta_i$), radians for $\phi_i$, and $10^{-6}$ inverse square degrees for $C_i$.}  
\label{Table2}
\end{table}

The two levels of the one-dimensional binary reflection grating -- separated by a height $T$ and thereby a relative return phase of $e^{i 4\pi T/\lambda}$ -- lead to interference of the electric field. We note that this simple model doesn’t include the full physics, e.g.\ phase shifts from the side walls of the binary grating, but this may be included in the $\phi$ fit parameter below. The interference is readily modelled using phasors \cite{McGilligan2016,Cotter2016} and shows a depth-dependent intensity \emph{behavior} in the first and zeroth diffracted orders with a form $A+B\sin(\phi+4\pi T/\lambda).$ The phase shift $\phi$, is expected to differ by $\pi$ between first and zeroth orders, with constants $A$ and $B$ approximately equal for strong zeroth-first order efficiency variation. By combining the diffraction angle and depth \emph{behavior}, we arrive at a diffraction efficiency fit of the form: 
\begin{equation}
\eta_i=(1 - C_i\, \theta^2)(A_i+B_i\sin(\phi_i+4\pi T/\lambda)), 
\label{fiteq}\end{equation} where the subscript $i$ can be 1, t or 0 depending on whether it refers to the 1$^\textrm{st}$, (t)otal or 0$^\textrm{th}$ order efficiency, respectively. 
To put this in context with our previous work  \cite{McGilligan2016}, we used a similar form of model but with $C=0,$ $\phi=\pi/2$, and for $\eta_1: A_1=-B_1,$ whilst for $\eta_\textrm{t}: B_\textrm{t}=0.$ Almost all data in Ref.~\cite{McGilligan2016} considered only a fixed relative etch depth to wavelength ratio of $T/\lambda\approx0.25$.

\begin{figure}[!b]
	\begin{minipage}{.49\textwidth}	\includegraphics[width=.8\columnwidth]{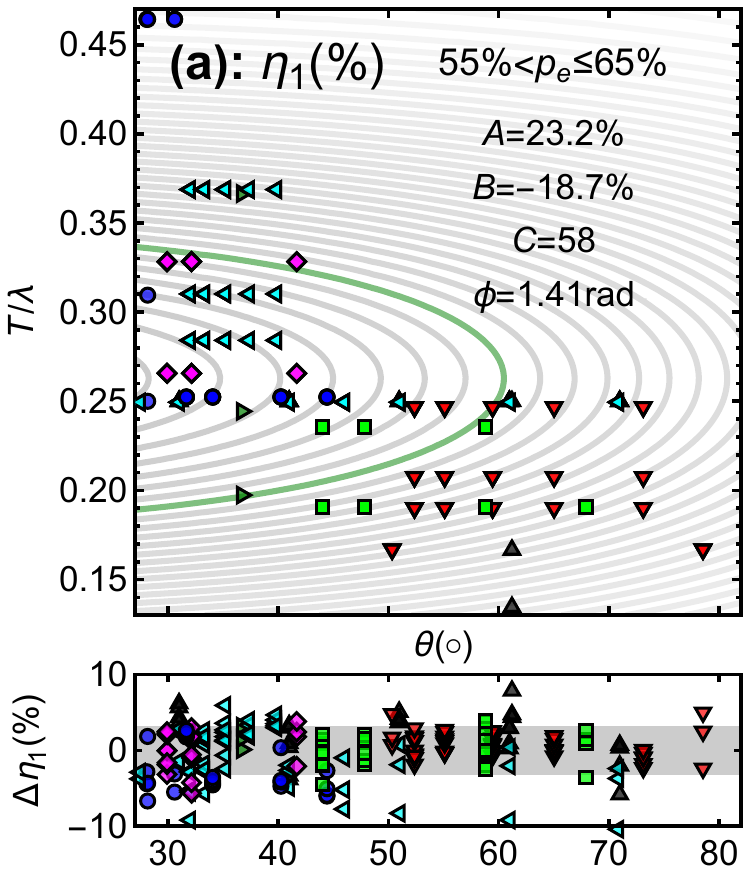}
		\makebox[0pt][c]{\raisebox{.85em}{\includegraphics[width=.2365\columnwidth]{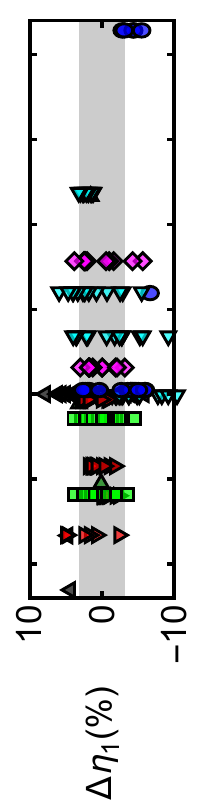}}\hspace*{-3.3em}}
\end{minipage}
	\begin{minipage}{.49\textwidth}
\includegraphics[width=.8\columnwidth]{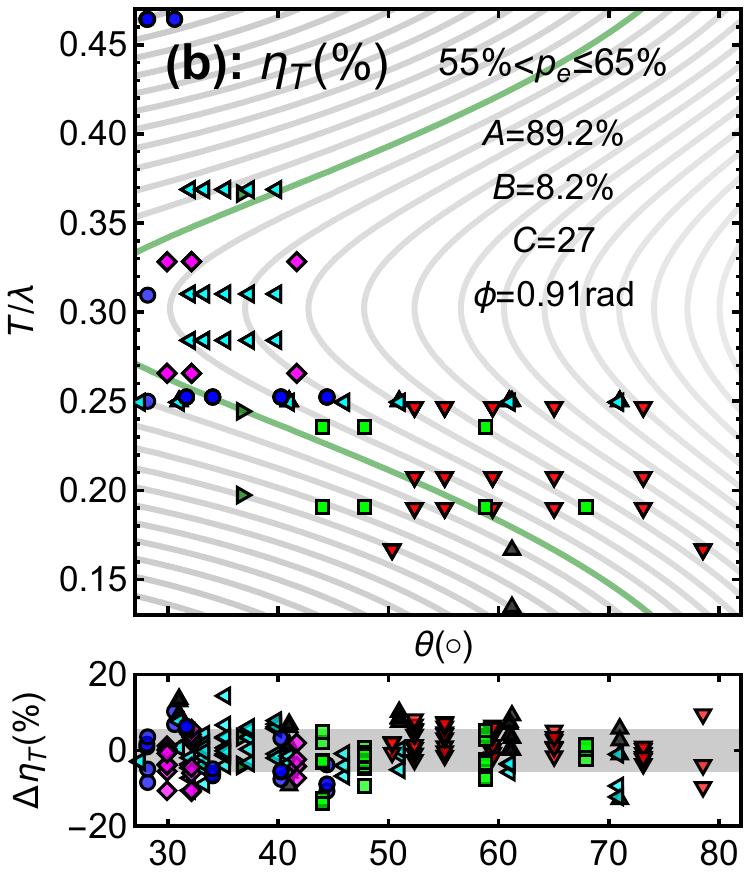}
		\makebox[0pt][c]{\raisebox{.85em}{\includegraphics[width=.2365\columnwidth]{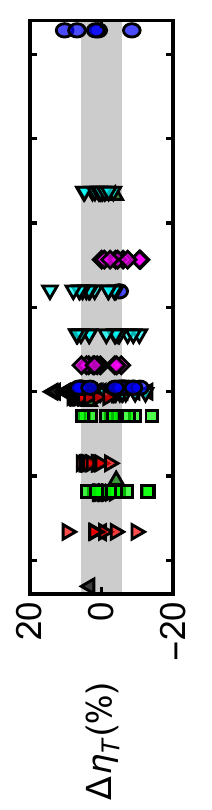}}\hspace*{-3.3em}}
\end{minipage}
	\caption{Contour plots show the  reflectivity-corrected fits to $55\%<p_\textrm{e}\leq65\%$ etch duty cycle data using Eq.~\ref{fiteq} for the first-order $\eta_1$ (a) and total diffracted $\eta_\textrm{t}$ (b) efficiencies, plotted vs.\ etch depth ratio $T/\lambda$ and diffraction angle $\theta=\arcsin(\lambda/d)$. Zeroth order efficiency can be inferred from  $\eta_0=\eta_\textrm{t}-2\eta_1$. Contours ranges from white to grey are (0-50)\% and  (50-100)\%, respectively, with 1\% contour spacing and a dark green contour at $\eta_1=33\%$ and $\eta_\textrm{t}=80\%$. Data point locations (dots) at test wavelengths of $(399, 420, 461, 532, 556, 689, 780)\,$nm are shown in purple, blue, cyan, dark green, light green, red and black (diamonds, circles, left-triangles, right-triangles, squares, down-triangles and up-triangles), respectively. The side images show the corresponding data point fit function residues in \% (the units of the contour plots), $\Delta \eta_1$ (a) and $\Delta \eta_\textrm{t}$ (b), with light gray zones indicating $\pm 1$ standard deviation (from Table~\ref{Table2}).\label{Fig2}}  
\end{figure}

We collated our data into those from three etch duty ranges (denoting  etched area as a fraction of the total): 
$45\%< p_\textrm{e}\leq 55\%$, $55\%< p_\textrm{e}\leq 65\%$, and $65\%< p_\textrm{e}\leq 75\%$, consisting of 174, 203 and 124 different grating+laser combinations, respectively. For these three  datasets \st{we fit} the empirical model Eq.~\ref{fiteq} \emph{was fitted} to both experimental data for $\eta_1$ and $\eta_\textrm{t}$ (Table~\ref{Table2}), yielding experiment-to-fit root-mean-square (RMS) residues to the $\eta_i$ datasets which are relatively small i.e.\ $\Delta\eta_1<5\%$ and $\Delta\eta_\textrm{t}<7\%$. Fitting to $\eta_0$ directly using Eq.~\ref{fiteq} led to slightly worse RMS errors to experiment than inferring the values from the two other fits via $\eta_0=\eta_\textrm{t}-2\eta_1$. We note that the simpler Ref.~\cite{McGilligan2016} grating model  would increase the Table~\ref{Table1} average fit errors for $\Delta\eta_1$, $\Delta\eta_\textrm{t}$, and $\Delta\eta_0,$ by $0.5\,\%$, $0.5\,\%$, and $1.7\,\%$, respectively -- most strongly affecting the zeroth order. Further details on the residuals are provided in the Supplementary Material.

From Table~\ref{Table2} the gratings with etch duty cycle range centred at $p_\textrm{e}=60\%$ \emph{generally have the best fit, overall diffraction efficiency, and strong first-order diffraction efficiencies (radial trapping and cooling forces)}, which matches our findings in Ref.~\cite{McGilligan2016}. For the remainder of the paper we will therefore only consider this optimal $p_\textrm{e}$ value. We illustrate the $\eta_1$ and $\eta_\textrm{t}$ fits and their residuals for the duty range $55\%< p_\textrm{e}\leq 65\%$ in Fig.~\ref{Fig2}, with the graphs pertaining to etch duties $45\%< p_\textrm{e}\leq 55\%$ and $65\%< p_\textrm{e}\leq 75\%$ in Fig.~S1 of  Supplement 1 \cite{dataset}. 

In addition we can also use Table~\ref{Table2} to infer other key parameters, e.g.\ the zeroth order diffraction efficiency $\eta_0$ as well as the `balance' parameters. 
Whilst radial intensity balance in a GMOT is achieved with a well-centred beam normally incident to the GMOT optic, the axial force of the incoming beam also needs to be balanced by the axial forces from the  diffracted beams. This is particularly important for a good optical molasses. For a spatially-uniform input beam, the axial beam-intensity balance due to $N$ first-order diffracted beams from a grating with $N$ sectors of 1D grating with wavelength-dependent coating reflectivity $R(\lambda)$ \cite{weber2002handbook} can be \emph{parametrized} by: \begin{equation}\label{eqbal}
\eta_{N\textrm{B}} = \frac{N R\, \eta_1}{1-R\,\eta_0}\;\;\;\textrm{i.e.}\;\;\;\eta_{3\textrm{B}} = \frac{3 R\, \eta_1}{1-R\,\eta_0}\;\textrm{(`Tri' grating),}\;\;\;\eta_{4\textrm{B}} = \frac{4 R\, \eta_1}{1-R\,\eta_0}\; \textrm{(`Quad' grating)},
\end{equation} where perfect axial beam-intensity is achieved at unity (100\%). We use the notation `Tri' and `Quad' to describe GMOT optics formed of $N = 3$ and $N=4$ sectors of linear binary grating, with the former having a slightly larger capture volume for more atoms, albeit with potential loading asymmetry.

Gratings \emph{containing holes with diameters greater than or equal to the MOT radius can} have no zeroth order `reflection' at the MOT location \emph{\cite{Vangeleyn2010}, with the additional benefits of allowing atomic loading \cite{Barker2019} and extraction \cite{Franssen2019}, as well as optical interrogation \cite{Bregazzi2021} and manipulation \cite{Lewis2022}}. The balance parameters describing gratings with holes (extra subscript H) are:
\begin{equation}\label{eqbalH}
\eta_{N\textrm{BH}} = N R\, \eta_1\;\;\;\textrm{i.e.}\;\;\;\eta_{3\textrm{BH}} = 3 R\, \eta_1\;\textrm{(`Tri' with hole),}\;\;\;\eta_{4\textrm{BH}} = 4 R\, \eta_1\;\textrm{(`Quad' with hole)}.\end{equation}

\section{Optimal bichromatic gratings}
\label{results2}

The results from section \ref{results1} can now be used to design a suitable binary diffraction grating for a given atomic species. Here, we present GMOT solutions for cooling the specific examples of Sr and Yb, which are often used in optical lattice clocks \cite{Takamoto2020,Fasano21}. The first step is to correct the model developed in section \ref{results1} for the metal reflectivity used at both `red' and `blue' wavelengths. Aluminium has high reflectivity $(>86\%)$ across wavelengths $(100 - 5000)\,$nm, and a flat response $R=(92.0\pm 0.5)\%$ over the broad optical wavelength range $(300-620)\,$nm \cite{weber2002handbook}. Al is also inert to many atomic species, and protected by a thin natural oxide layer. For both Sr and Yb gratings we therefore mainly consider aluminium coatings, noting that other metal coatings may also be suitable -- we have \emph{realized} Rb GMOTs with Al, Au, Pd  and Pt as grating coatings.

\begin{figure}[!t]
\centering
\includegraphics[width=1\columnwidth]{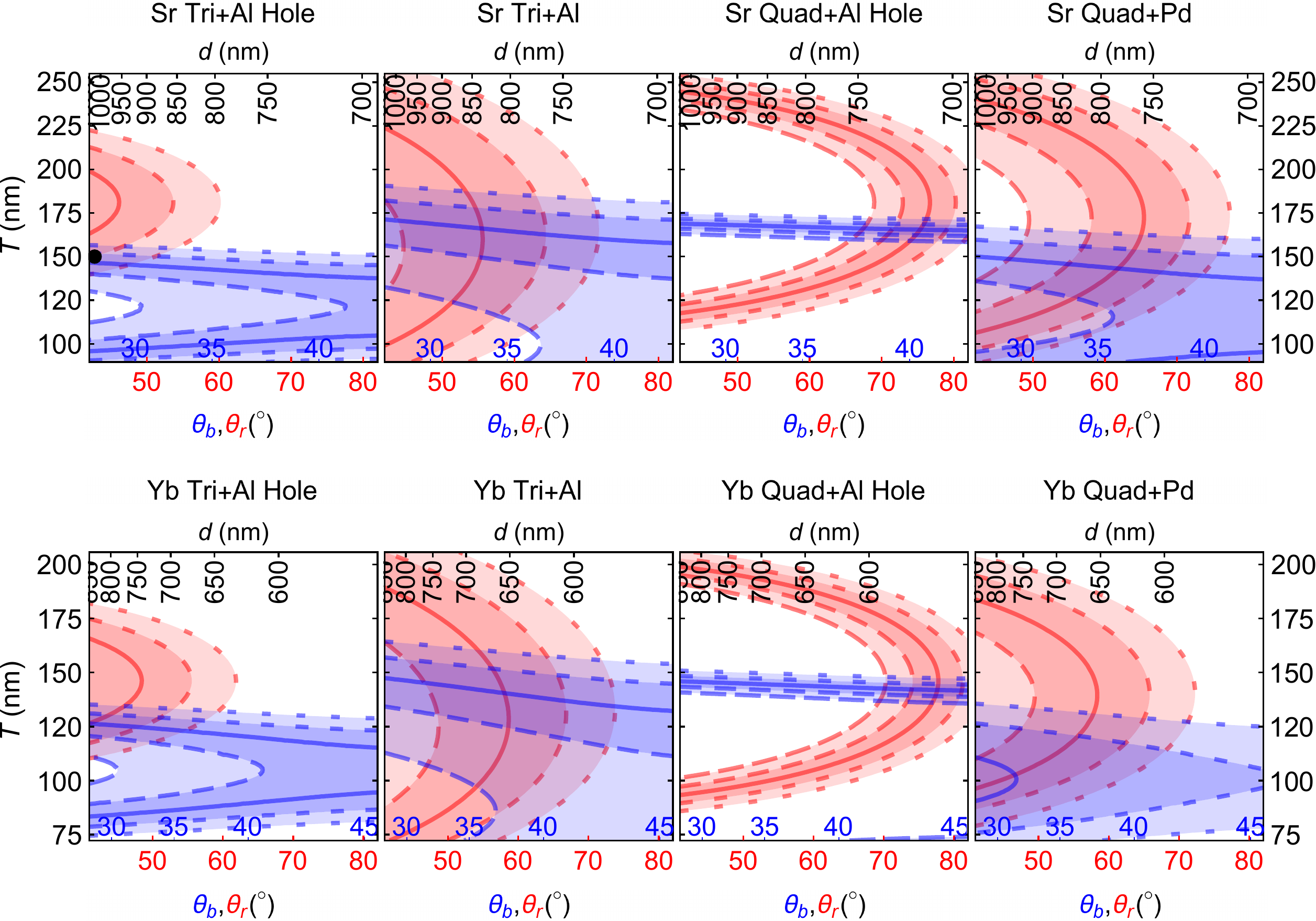}
\caption{Grating balance parameters (Eqs.~\ref{eqbal},\ref{eqbalH}) for Sr (upper row) and Yb (lower row) at both red $\lambda_\textrm{r}$ and blue $\lambda_\textrm{b}$ wavelengths as a function of binary grating parameter space (period $d$ vs.\ etch depth $T$). The GMOT geometries  considered (left to right) are: Al-coated Tri-with-hole  ($\eta_{3\textrm{BH}}$, Sr black dot is from  \cite{Sitaram2020,elgee22}), Tri  ($\eta_{3\textrm{B}}$), Quad-with-hole ($\eta_{4\textrm{BH}}$); as well as Pd-coated Quad ($\eta_{4\textrm{H}}$).  Contours and shaded regions are displayed in their respective red and blue colors, denoting wavelength. Contours are drawn at balances $(90, 95, 100, 105,110)\%$ with increasing dash width, excepting the solid 100 \% balance contour. The $(95-105)\%$ and $(90-110)\%$ regions are shaded dark and light, respectively.\label{Fig3}} 
\end{figure}

Using Eq.~\ref{fiteq} and the optimal $55\%<p_e\leq 65\%$ duty fit results from Table \ref{Table2}, the reflectivity-adjusted balance (Eq.~\ref{eqbal}) can be calculated for both wavelengths and for different grating architectures.  We considered the balance parameter for four macroscopic grating designs, namely three-sector (Tri) optics with and without holes at the grating centre ($\eta_{3\textrm{BH}}$, $\eta_{3\textrm{B}}$), and the equivalent for Quad optics  ($\eta_{4\textrm{BH}}$, $\eta_{4\textrm{B}}$), shown in Fig.~\ref{Fig3} left to right, respectively, with Al, Al, Al, Pd coatings. 
The results in Fig.~\ref{Fig3} are given for both Sr (upper row) and Yb (lower row). 
The grating period considered is restricted to $\lambda_\textrm{r} < d < 2 \lambda_\textrm{b}$ to ensure only zeroth-order \textit{and} first-order diffraction occurs at both wavelengths (i.e. $30^\circ<\theta_\textrm{r,b}<90^\circ)$. The plots use an x-axis of the `red' diffraction angle $\theta_\textrm{r}$ to aid comparison to Fig.~\ref{Fig2}, and we restrict to $\theta_\textrm{r}<80^\circ$ for which experimental grating data exists.

In Fig.~\ref{Fig3}, we now consider the GMOT geometries left-to-right. For the Tri with hole the model indicates $\theta_\textrm{b}<30^\circ$ may be suitable at both wavelengths, matching the Al-coated Sr Tri grating with hole of Ref.~\cite{Sitaram2020,elgee22} (black dot). We note our model is likely still valid a little below $\theta_\textrm{b}=30^\circ$, if the large-angle second-order blue diffraction is minimal. For both atomic species with holeless Al-coated Tri designs we see a large overlap of good balance in grating parameter space, suggesting a promising area for future GMOT experiments.   
In Quad Al-coated gratings with holes the overlap is much smaller than for Tri gratings, but the contour intersection still indicates a more specific useful grating region. This region has experimental data for the red and blue (Fig.~\ref{Fig2}), albeit at high red diffraction angle and hence low red GMOT overlap height.  
To \emph{realize}  regular Quad grating GMOTs, or to reduce the extremity of the `red' diffraction angle for quads with holes, an alternative is to pick a different metal coating, here we show Pd instead of Al for the Quads, which greatly extends their useful 
parameter space. 
A functional Sr Quad with hole GMOT has already been experimentally demonstrated \cite{bondza22}, albeit with a multi-layer metal coating, giving reduced blue reflectivity. This difference prevents direct comparison, but matches expectations if the balance parameters in Fig.~\ref{Fig3} were adjusted accordingly.

The intersection of $100\%$ balance contours in Fig.~\ref{Fig3} would na\"{i}vely lead to the optimal GMOT grating, however we now stress caveats to this approach. The two experiments with bi-chromatic Sr gratings \cite{bondza22} and  \cite{Sitaram2020,elgee22}, have $\lambda_\textrm{r}$ balances of $72\%$ and $111\%$, respectively, demonstrating that even the narrow-linewidth red cooling works in a large range of balances, which matches our experience with Rb GMOT balance \cite{Nshii2013,Burrow2021}. This implies that balances constrained to $(100\pm10)\%$ are in fact overly restrictive and it is likely a wider range of grating parameter space is available. 

Seo \textit{et al}.\cite{Seo2021} also reported on a six-parameter simulation of GMOTs, \emph{optimized} by machine learning, that indicated higher optical balances capture more atoms. 
Caution must be used however, as there is a diffraction-angle specific value corresponding to a level of optical balance which precludes MOT formation \cite{Vangeleyn2009,vangeleynthesis,bondza22,newpaper}. As a technically simple fix for GMOT under-balance, a disk of weak neutral density (ND) filter in the center of the GMOT input laser beam can also be used \cite{McGilligan2015}. Conversely a weak ND filter with a disk cut from its centre can restore optical balance when there is too much diffracted power.  In principle, bespoke filters could be fabricated to tune the balance of \textit{both} wavelengths to a designated level.

Other considerations for GMOT grating design include the input beam-profile used \cite{McGilligan2015} -- so far we have assumed a grating illuminated with a spatially uniform beam. Without beam-shaping, most input beams have a Gaussian transverse intensity profile $I_0 \exp(-2 r^2/w^2)$ with beam waist $w$, leading to a balance parameter that increases with distance $z$ above the grating centre.
Specifically, this modifies any instance of $\eta_1$ in Eq.~\ref{eqbal} to $\eta_1'=\eta_1 \exp(-2(z\tan\theta)^2/w^2)$, where both the diffraction angle $\theta$ and $w$ depend on the MOT laser wavelength. 
Typically with alkali metal atoms, over-expanding the beam simplifies this consideration, but the primary cooling transitions for alkaline-earth-like atoms have a large saturation intensity, at wavelengths where relatively high power laser sources can be complex or expensive.

Another vital design consideration for GMOTs is the trade-off between axial and radial trapping and cooling forces, as well as their local and spatially averaged (trap depth) maxima. These compromises strongly depend on grating diffraction angle \cite{Vangeleyn2009, vangeleynthesis,Vangeleyn2010,Lee2013, McGilligan2015,eckel2023,newpaper}. There is evidence that \emph{vapor}-loaded Rb GMOTs have largest atom number at diffraction angles around 45$^\circ$ \cite{McGilligan2015,Seo2021}, which has been corroborated theoretically \cite{Seo2021,newpaper}. It is therefore also worth considering \emph{favoring} the primary cooling transition for loading in the case of Sr and Yb, and $\theta_\textrm{b}\approx 45^\circ$ corresponds to the far right of all images in Fig.~\ref{Fig3}. 
Furthermore most work to date considers only simple atomic models and neglects internal state dynamics, which strongly affect sub-Doppler cooling mechanisms \cite{Landini11,Barker2022}.

Our empirical grating model, in conjunction with a wavelength-specific coating reflectivity, gives a balance parameter that allows one to simply determine the optical properties of  a given bichromatic binary grating. However, several other factors -- including the effects of the multi-level structure of real atoms on GMOTs \cite{eckel2023} -- may need consideration to tailor a grating to a given user's GMOT design criteria, and we plan to investigate further \cite{newpaper}.

\section{Conclusion}

By processing data from a wide variety of diffraction grating-wavelength combinations we arrived at a simple dimensionless empirical model of diffraction efficiencies. The model, in conjunction with coating reflectivity, allows one to tailor grating design to form compromise fabrication solutions that works for laser-cooling and trapping in GMOTs at two wavelengths. 
The grating model could also be used to design single-color GMOTs more accurately, or extended to an optically simpler three-color GMOT cooling method for group-III elements, like Indium \cite{Yu2022}.   

The multi-wavelength diffractive optics proposed here will be ideal for any high-end application where the size, weight and power of the laser cooling package is critical, particularly for earth- and space-based portable quantum technologies \cite{McGilligan2022,Abend2023}.

\begin{backmatter}
\bmsection{Acknowledgments}
We thank James McGilligan, William McGehee and Roger Brown for thorough proofreading; and James McGilligan, David Burt and Brendan Casey for valuable conversations.

\bmsection{Funding}
InnovateUK+EPSRC projects \href{https://gow.epsrc.ukri.org/NGBOViewGrant.aspx?GrantRef=EP/M013294/1}{EP/M013294/1}, \href{https://gow.epsrc.ukri.org/NGBOViewGrant.aspx?GrantRef=EP/M50824X/1} {EP/M50824X/1}, 
\href{https://gow.epsrc.ukri.org/NGBOViewGrant.aspx?GrantRef=EP/R002371/1}{EP/R002371/1}, 
\href{https://gow.epsrc.ukri.org/NGBOViewGrant.aspx?GrantRef=EP/T001046/1}{EP/T001046/1}. 

\bmsection{Disclosures}
The authors declare no conflicts of interest.

\bmsection{Data availability} Data underlying the results presented in this paper are available in the Dataset, Ref.~ \cite{dataset}. For the purpose of open access, the author(s) has applied a Creative Commons Attribution (CC BY) licence to any Author Accepted Manuscript (AAM) version arising from this submission.

\bmsection{Supplemental document}
See Supplement 1 for supporting content, chiefly an extended model error analysis via figures S1-S4 which determine whether the duty cycle range, test laser wavelength, diffraction angle, or etch depth show any biases -- with nothing significant observed. A discussion of residual errors from measuring gratings with nominally the same manufacturing parameters is also provided.  

\end{backmatter}

%\bibliography{Biblio}

\end{document}